\documentclass[prl,twocolumn,showpacs,preprintnumbers,amsmath,amssymb,superscriptaddress, bibnotes]{revtex4}

\usepackage{color}
\usepackage{graphicx}
\usepackage{dcolumn}
\usepackage{bm}

\begin{document}
\newcommand{\red}[1]{\textcolor{red}{#1}}

\title{Evidence for the presence of the Fulde-Ferrell-Larkin-Ovchinnikov state in CeCu$_2$Si$_2$ revealed using $^{63}$Cu NMR }

\author{Shunsaku~Kitagawa}
\email{kitagawa.shunsaku.8u@kyoto-u.ac.jp}
\author{Genki~Nakamine}
\author{Kenji~Ishida}
\affiliation{Department of Physics, Kyoto University, Kyoto 606-8502, Japan}
\author{H.~S.~Jeevan}
\author{C.~Geibel}
\author{F. Steglich}
\affiliation{Max-Planck Institute for Chemical Physics of Solids, D-01187 Dresden, Germany}

\date{\today}

\begin{abstract}
Nuclear magnetic resonance  measurements were performed on CeCu$_{2}$Si$_{2}$ in the presence of a magnetic field close to the upper critical field $\mu_{0} H_{\rm c2}$ in order to investigate its superconducting (SC) properties near pair-breaking fields. 
In lower fields, the Knight shift and nuclear spin-lattice relaxation rate divided by temperature $1/T_1T$ abruptly decreased below the SC transition temperature $T_{\rm c}(H)$, a phenomenon understood within the framework of conventional spin-singlet superconductivity.
In contrast, $1/T_1T$ was enhanced just below $T_{\rm c}(H)$ and exhibited a broad maximum when magnetic fields close to $\mu_0H_{\rm c2}(0)$ were applied parallel or perpendicular to the $c$ axis, although the Knight shift decreased just below $T_{\rm c}(H)$.
This enhancement of $1/T_1T$, which was recently observed in the organic superconductor $\kappa$-(BEDT-TTF)$_{2}$Cu(NCS)$_{2}$, suggests the presence of high-density Andreev bound states in the inhomogeneous SC region, a hallmark of the Fulde-Ferrell-Larkin-Ovchinnikov phase.
\end{abstract}

\maketitle
The Fulde-Ferrell-Larkin-Ovchinnikov (FFLO) state, predicted half a century ago\cite{P.Fulde_PR_1964, A.I.Larkin_JETP_1965}, is one of the exotic superconducting (SC) states that has not been fully characterized.
There are two well-known pair-breaking mechanisms exhibited by a type-II superconductor under a magnetic field.
One is the orbital pair-breaking effect related to the emergence of Abrikosov vortices, in which superconductivity is destroyed at the vortex cores.
The other is the Pauli pair-breaking effect, which originates from the Zeeman effect produced by the presence of external fields.
When the Zeeman-splitting energy is as high as the condensation energy of superconductivity, in principle, superconductivity becomes unstable and transitions to the normal state with first-order character.
Realization of the FFLO state would be expected in the vicinity of the upper critical magnetic field ($H_{\rm c2}$) when the Pauli pair-breaking effect predominates over the orbital pair-breaking effect.

In the FFLO state, the spin-singlet Cooper pair is formed between spin-split Fermi surfaces; thus, the Cooper pairs have finite center-of-mass momenta.
Consequently, the spatially modulated superconducting state is realized.
Since the orbital-limiting field in a single band is expressed as, $H_{\rm c2}^{\rm orb} = \phi / 2 \pi \xi^2$ with $\xi = \hbar v_{\rm F}/\Delta_0 \pi$ and $v_{\rm F} = \hbar k_{\rm F} /m^*$ \cite{R.R.Hake_APL_1967}, where $k_{\rm F}$, $\Delta_{0}$ and $m^*$ are the Fermi wave vector, the superconducting gap, and the effective mass of electron, respectively, a heavy-electron mass is important for the realization of the FFLO state.
Therefore, $H_{\rm c2}^{\rm orb}$ becomes large and the Pauli pair-breaking effect predominates in heavy-fermion superconductors\cite{K.Gloos_PRL_1993,A.Bianchi_PRL_2003,Y.Matsuda_JPSJ_2007}; and quasi-two-dimensional organic superconductors in a magnetic field parallel to the conducting layers\cite{J.Singleton_JPCS_2000,R.Lortz_PRL_2007}.

Mayaffre {\it et al}. recently performed high-field nuclear magnetic resonance (NMR) measurements for a magnetic field parallel to the conducting plane, in $\kappa$-(BEDT-TTF)$_{2}$Cu(NCS)$_{2}$, which is one of the candidates for realizing the FFLO state, and they found a clear enhancement of the nuclear spin-lattice relaxation rate $1/T_1$ in the field range between the Pauli-limiting field $H_{\rm P}$ and $H_{\rm c2}(T)$\cite{H.Mayaffre_NatPhys_2014}, indicated by the appearance of the sharp bound states located away from $E_{\rm F}$.
The enhancement of $1/T_1$ suggested the formation of FFLO state, and it is in good agreement with the $1/T_1$ calculation\cite{B.M.Rosemeyer_PRB_2016}.
Therefore, taking $1/T_1$ measurements around $H_{\rm c2}$ is recognized as a valuable method for searching the FFLO state.

CeCu$_{2}$Si$_{2}$ was the first heavy fermion superconductor to be studied, with $T_{\rm c}$ = 0.6~K\cite{F.Steglich_PRL_1979}, crystallized in the tetragonal ThCr$_2$Si$_2$-type structure and space group $I4/mmm$ ($D_{4h}^{17}$, No.139).
After the discovery of superconductivity, CeCu$_2$Si$_2$ was considered to be a nodal unconventional superconductor based on experiments on polycrystalline samples\cite{Y.Kitaoka_JPSJ_1986,K.Ishida_PRL_1999,F.Steglich_PhysicaB_1996,K.Fujiwara_JPSJ_2008}.
However, recent specific-heat $C_e$\cite{S.Kittaka_PRL_2014} and thermal conductivity measurements\cite{T.Yamashita_SciAdv_2017} on an SC phase dominant ($S$-type) single crystal of CeCu$_2$Si$_2$ down to a temperature of 40 mK strongly suggest that CeCu$_2$Si$_2$ possesses two BCS-type gaps, the magnitudes of which differ greatly.
The multiband full-gap behavior was also confirmed by our recent nuclear magnetic resonance (NMR) measurements on an $S$-type single-crystal sample\cite{S.Kitagawa_PRB_2017}.
Although a lot of experimental and theoretical works have been performed in order to clarify the SC-gap symmetry\cite{S.Kittaka_PRB_2016,S.Kitagawa_PRB_2017,T.Yamashita_SciAdv_2017,T.Takenaka_PRL_2017, G.M.Pang_PNAS_2018,Y.Li_PRL_2018}, it has not been settled yet.
At present, a full understanding of the SC-gap state is an important issue in the discussion of CeCu$_2$Si$_2$.

Considering the parity of the superconductivity, the decrease of the Knight shift in the SC state\cite{K.Ueda_JPSJ_1986} and the strong suppression of $\mu_0 H_{\rm c2}(T)$ along any magnetic field direction indicates a spin-singlet pairing\cite{W.Assmus_PRL_1984,H.A.Vieyra_PRL_2011,S.Kittaka_PRB_2016}.
Furthermore, the unusual decrease of $\mu_0 H_{\rm c2}(T)$ below $\sim$ 0.2~K was observed by thermodynamic and resistivity measurements\cite{W.Assmus_PRL_1984,H.A.Vieyra_PRL_2011,S.Kittaka_PRB_2016}; thus the Pauli pair-breaking effect seems to dominate.
However, the first-order phase transition expected in such a case has not been clearly observed.
It was suggested that the multiband character of superconductivity may suppress a first-order transition\cite{S.Kittaka_PRB_2016}.
The conditions for the realization of the FFLO state appear to be satisfied by CeCu$_{2}$Si$_{2}$, but the possible presence of the FFLO state has so far been discarded due to the large residual resistivity, and no experimental evidence of the FFLO state has yet been reported.
This is partly due to limitations in the ability of experimental probes to detect the FFLO state.

In this study, we have performed $^{63}$Cu NMR measurements in order to investigate the SC properties of single-crystal CeCu$_2$Si$_2$ near $H_{\rm c2}$.
We found an anomalous enhancement of 1/$T_1T$ just below $\mu_{0} H_{\rm c2}(T)$ in the presence of a magnetic field parallel or perpendicular to the $c$ axis, $H \parallel c$ and $H \perp c$, respectively.
Comparing these results to those reported for $\kappa$-(BEDT-TTF)$_{2}$Cu(NCS)$_{2}$, our results suggest that the FFLO phase is formed in the narrow region just below $\mu_{0} H_{\rm c2}$ in CeCu$_{2}$Si$_{2}$.

Single crystals of CeCu$_{2}$Si$_{2}$ were grown by the flux method\cite{S.Serio_PSPB_2010}.
A high-quality $S$-type single crystal was used for the NMR measurements, which is the same as those used in previous NMR/nuclear quadrupole resonance measurements\cite{S.Kitagawa_PRB_2017}.
The field dependence of $T_{\rm c}(H)$ [$\mu_0H_{\rm c2}(T)$] was obtained by ac susceptibility measurements using an NMR coil.
A conventional spin-echo technique was used for the NMR measurements.
Low-temperature measurements below 1.7 K were carried out with a $^3$He - $^4$He dilution refrigerator, in which the sample was immersed in the $^3$He - $^4$He mixture in order to avoid radio frequency heating during measurements. 
The external fields were controlled by a single-axis rotator with an accuracy above 0.5$^{\rm o}$.
The $^{63}$Cu-NMR spectra (nuclear spin $I$ = 3/2, and nuclear gyromagnetic ratio $^{63}\gamma/2\pi = 11.285$~MHz/T) were obtained as a function of frequency in a fixed magnetic field, and the $^{63}$Cu Knight shift of the sample was calibrated using the $^{63}$Cu signals from the NMR coil.
The $^{63}$Cu nuclear spin-lattice relaxation rate $1/T_1$ was determined by fitting the time variation of the spin-echo intensity after saturation of the nuclear magnetization to a theoretical function for the central transition of $I$ = 3/2\cite{A.Narath_PR_1967,D.E.MacLaughlin_PRB_1971}.
\begin{figure}[!b]
\vspace*{10pt}
\begin{center}
\includegraphics[width=8.5cm,clip]{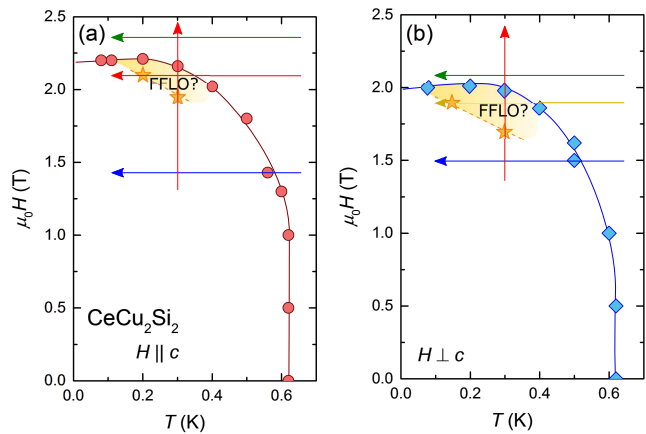}
\end{center}
\caption{Temperature dependence of $\mu_0 H_{\rm c2}$ for $H \parallel c$(a) and $H \perp c$(b) on CeCu$_2$Si$_2$ obtained by ac susceptibility measurements using an NMR coil.
Stars indicate the lowest magnetic field where $1/T_1T$ is larger than that in the normal state.
The solid curves and broken lines are provided as a guide for the eye.
Arrows indicate the magnetic field and temperature scans covered by the NMR measurements.
}
\label{Fig.1}
\end{figure}

\begin{figure*}[!tb]
\vspace*{10pt}
\begin{center}
\includegraphics[width=14cm,clip]{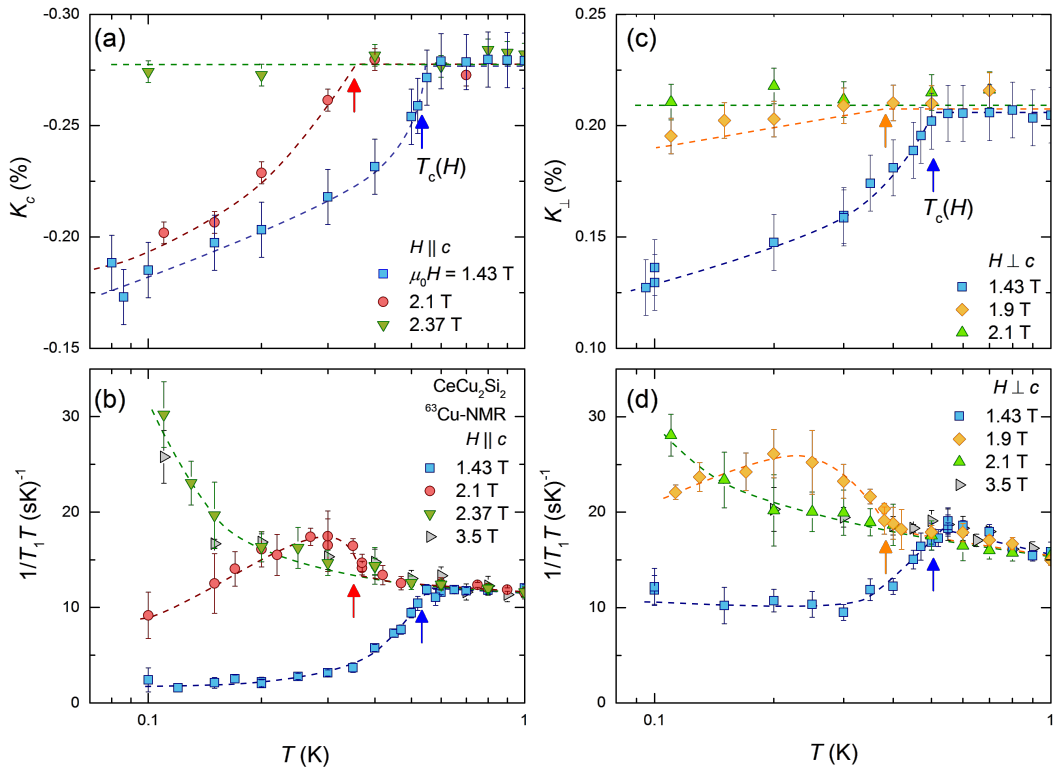}
\end{center}
\caption{Temperature dependence of the Knight shift $K(T)$ and $1/T_1T$ under various magnetic fields for $H \parallel c$ [(a): $K(T)$,(c): $1/T_1T$] and $H \perp c$ [(b): $K(T)$,(d): $1/T_1T$].
The arrows indicate $T_{\rm c}(H)$.
The dashed curves are given as a guide for the eye.
}
\label{Fig.2}
\end{figure*}
\begin{figure}[!tb]
\vspace*{10pt}
\begin{center}
\includegraphics[width=8.5cm,clip]{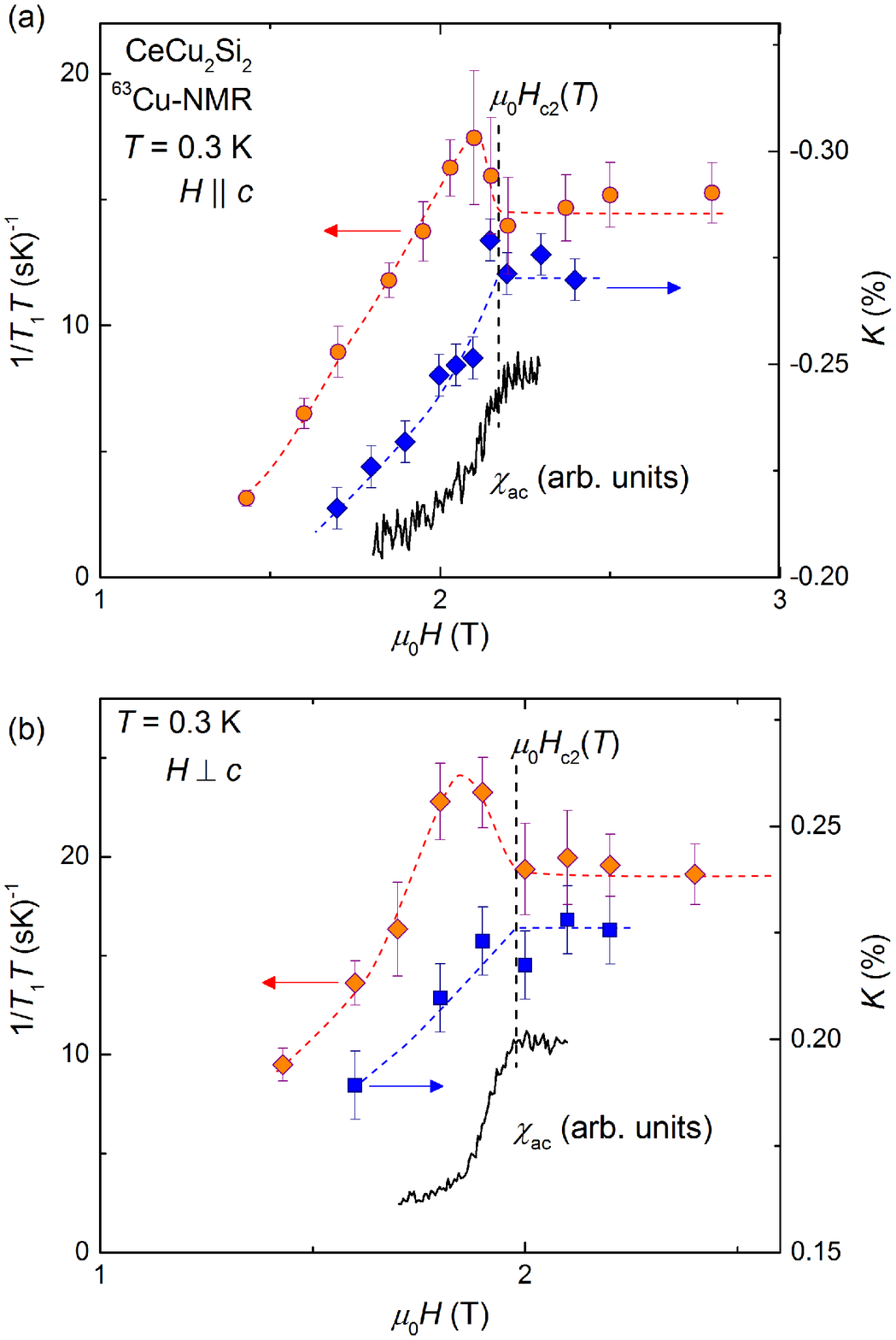}
\end{center}
\caption{Magnetic field dependence of $1/T_1T$ together with the Knight shift and $\chi_{ac}$ measured using an NMR coil at 0.3~K for $H \parallel c$(a) and $H \perp c$(b).
The broken lines indicate $\mu_0H_{\rm c2}(0.3~{\rm K})$.
The dashed curves are given as guides for the eye.
}
\label{Fig.3}
\end{figure}
SC phase diagrams were obtained by ac susceptibility measurements, as shown in Fig.~\ref{Fig.1}.
The strong suppression of $\mu_0 H_{\rm c2}(T)$ and the decrease of $\mu_0 H_{\rm c2}(T)$ below $\sim$0.2~K were observed, in good agreement with previous thermodynamic and resistivity measurements\cite{W.Assmus_PRL_1984,H.A.Vieyra_PRL_2011,S.Kittaka_PRB_2016}. 
Furthermore, the Pauli-limiting fields $H_{\rm P}$(0), a field at which SC condensation energy is equal to Zeeman energy, estimated from the formula 
\begin{align}
H_{\rm P} = H_{c}/\sqrt{4\pi~\delta\chi} \hspace*{0.5cm} \textrm{(in cgs units)}
\end{align}
are $\mu_0 H_{\rm P}$ = 2.0 T and 1.56 T for $H \parallel c$ and $H \perp c$, respectively, which are close to the experimental values shown in Fig.~\ref{Fig.1}.
Here, the SC critical field $H_c$ is 620 Oe \cite{S.Kittaka_PRB_2016}, and $\delta \chi$ is the difference in the susceptibility between the normal and SC state, $\delta \chi = \chi_{\rm N} - \chi_{\rm SC}$, which is estimated using the relation $\delta \chi = \left( \mu_{\rm B} N_{\rm A}/^{63}A_{\rm hf}\right)\delta K$.
The experimental values for the hyperfine coupling constant of $^{63}A_{\rm hf}$ and the decrease of the Knight shift $\delta K$ of $^{63}$Cu below $T_{\rm c}$ are used\cite{H.Tou_JPSJ_2005}.
In addition, the orbital-limiting fields $\mu_0 H_{\rm c2}^{\rm orb}$ obtained from the initial slope of $H_{\rm c2}$, $H_{\rm c2}^{\rm orb} \simeq - 0.7T_{\rm c}dH_{\rm c2}/dT |_{T=T_{c}}$, are 14.5~T and 10~T for $H \parallel c$ and $H \perp c$, respectively\cite{S.Kittaka_PRB_2016}; thus the Maki parameter $\alpha =\sqrt{2} H_{\rm c2}^{\rm orb}/H_{\rm P}$\cite{K.Maki_PR_1966} is estimated to be 9.5 and 7.1 for $H \parallel c$ and $H \perp c$, respectively assuming $H_{\rm P} = H_{\rm c2}(0)$.  
These values are much larger than the required minimum value of $\alpha$ for the formation of the FFLO state $\alpha = 1.8$\cite{L.W.Gruenberg_PRL_1966}, and even larger than $\alpha$ for CeCoIn$_5$ ($\alpha_{\parallel c} = 5.0$ and $\alpha_{\perp c} = 4.6$)\cite{K.Kumagai_PRL_2006}, for which the novel magnetic state related with the FFLO state was observed.
This larger $\alpha$ value appears to be favorable for the formation of the FFLO state near $H_{\rm c2}$. 
       
Figure~\ref{Fig.2} shows the temperature dependence of the Knight shift $K(T)$ and $1/T_1T$ in the presence of various magnetic fields for $H \parallel c$ and $H \perp c$.
In the case of $H \parallel c$, $K(T)$ and $1/T_1T$ clearly decreased below $T_{\rm c}(H)$ at 1.43~T as has been previously reported\cite{S.Kitagawa_PRB_2017}, which is consistent with the behavior in a conventional spin-singlet superconductor.
In contrast, $1/T_1T$ was enhanced just below $T_{\rm c}(H)$ and shows a broad maximum around 0.3~K at 2.1~T, close to $\mu_0H_{\rm c2}(0)$.
However, $K(T)$ decreased just below $T_{\rm c}(H)$ and did not exhibit a similar enhancement.
In the presence of a higher field of 2.37~T [$>\mu_0H_{\rm c2}(0)$], $1/T_1T$ continuously increased as the temperature decreased to a value of 120~mK, suggestive of a nearby AFM quantum critical point, and no clear magnetic field dependence was observed above $\mu_0H_{\rm c2}(0)$ up to 3.5~T.
Comparing the results obtained at 2.1~T with those above $H_{\rm c2}$, it is apparent that the enhancement of $1/T_1T$  was observed in the narrow field just below $\mu_0 H_{\rm c2}(T)$, and that the enhanced $1/T_1T$ was larger than that of the normal state $1/T_1T$.
A similar enhancement of $1/T_1T$ was also observed for $H \perp c$, but the enhancement of $1/T_1T$ in this case was much clearer than that for $H \parallel c$, probably due to the larger residual density of states as $T \rightarrow 0$.
The clear enhancement of $1/T_1T$ observed at 2,1~T for $H \parallel c$ and 1.9~T for $H \perp c$ cannot be simply interpreted by the competition between antiferromagnetic fluctuations and superconductivity, since the occurrence of superconductivity always suppresses $1/T_1T$ due to the opening of the SC gap. 
Alternatively, the enhancement of $1/T_1T$ clearly indicates the appearance of the high density of states only near $\mu_0 H_{\rm c2}$.

The enhancement of $1/T_1T$ was also indicated by the magnetic field dependence of $1/T_1T$ measured at 0.3~K as shown in Fig.~\ref{Fig.3}.
Above $\mu_0H_{\rm c2}(T)$, $K$ and $1/T_1T$ are independent of the magnetic field.
While $K$ monotonically decreased with the decreasing field below $\mu_0H_{\rm c2}(T)$, as determined by $\chi_{\rm ac}$ due to the formation of spin-singlet Cooper pairs, $1/T_1T$ exhibited a broad maximum below $\mu_0H_{\rm c2}(T)$ for both magnetic-field directions.
Since the phase transition between the FFLO and SC phases was not observed in $1/T_1T$, we plot the lowest magnetic field where $1/T_1T$ is larger than that in the normal state as the transition temperature as shown in Fig.~\ref{Fig.1}.
It is noted that we did not observe any clear anomaly in the field dependence of linewidth at $\mu_0H_{\rm c2}(T)$ in CeCu$_{2}$Si$_{2}$, although it gives the information about the inhomogeneity of SC state\cite{H.Mayaffre_NatPhys_2014}.
Furthermore, it seems that the FFLO state is not the ground state at $T = 0$~K in Fig.~\ref{Fig.1}, which is different from both of the FFLO states in $\kappa$-(BEDT-TTF)$_2$Cu(NCS)$_2$\cite{H.Mayaffre_NatPhys_2014} and CeCoIn$_5$\cite{Y.Matsuda_JPSJ_2007}.
This might originate from the experimental technique.
In NMR measurements, strong AFM fluctuations make it difficult the observation of the anomaly at low temperatures.
To clarify the boundary between the SC and FFLO phase, other experimental techniques are necessary.

Here, we compare the NMR results obtained using CeCu$_{2}$Si$_{2}$ with those of $\kappa$-(BEDT-TTF)$_{2}$Cu(NCS)$_{2}$\cite{H.Mayaffre_NatPhys_2014}.
While both compounds exhibit a clear enhancement of $1/T_1T$ just below $\mu_0H_{\rm c2}(T)$, there are several differences between these two compounds.
First, the normalized increase of $1/T_1T$, $(1/T_1T)^{\rm peak}/(1/T_1T)^{\rm normal}$, differs.
Although $1/T_1T$ increased almost twofold in the case of $\kappa$-(BEDT-TTF)$_{2}$Cu(NCS)$_{2}$, in the present case $(1/T_1T)^{\rm peak}/(1/T_1T)^{\rm normal}$ estimated from the magnetic field dependence of $1/T_1T$ is $\sim 1.2$ for $H \parallel c$ and $\sim 1.3$ for $H \perp c$.
The increase in ratio of $1/T_1T$ depends on the SC gap symmetry and the details of the Fermi surfaces of the systems\cite{B.M.Rosemeyer_PRB_2016}.
Second, the enhancement of $1/T_1T$ was observed for $H \parallel c$ and $H \perp c$ in CeCu$_{2}$Si$_{2}$ while this enhancement was only observed for $H \parallel$ conducting plane in the case of $\kappa$-(BEDT-TTF)$_{2}$Cu(NCS)$_{2}$.
This difference originates from the dimensionality of the systems.
$\kappa$-(BEDT-TTF)$_{2}$Cu(NCS)$_{2}$ is a quasi-two-dimensional superconductor; in this case, $\mu_0 H_{\rm c2}^{\rm orb}$ is high only when the magnetic field is applied parallel to the conducting plane.
On the other hand, CeCu$_{2}$Si$_{2}$ is a three-dimensional (3D) heavy fermion superconductor, and so the high $\mu_0 H_{\rm c2}^{\rm orb}$ is due to heavy electron mass and not related to the magnetic field directions.
Third, the $H-T$ phase diagram and the regions where the enhancement of $1/T_1T$ was observed differ.
In the case of $\kappa$-(BEDT-TTF)$_{2}$Cu(NCS)$_{2}$, $\mu_0H_{\rm c2}(T)$ exhibits an upturned behavior above $\mu_0 H_{\rm P} (0)$\cite{R.Lortz_PRL_2007,B.Bergk_PRB_2011,C.C.Agosta_PRB_2012} and the $1/T_1T$ enhancement was observed in this region.
On the other hand, $\mu_0H_{\rm c2}(T)$ is strongly suppressed and rather decreases below $\sim$0.2~K in the case of CeCu$_{2}$Si$_{2}$ where $1/T_1T$ enhancement was observed in a rather narrow region close to $\mu_0H_{\rm c2}(0)$.
According to theory, the FFLO region is narrower in a 3D superconductor than in a 2D superconductor\cite{Y.Matsuda_JPSJ_2007}, which is also consistent with the experimental results.

We also point out that the narrow region of the FFLO state in CeCu$_2$Si$_2$ can be explained with the impurity scattering effect. 
It was shown from theoretical studies that the phase boundary of the Pauli-limited upper critical field and of the FFLO state depends sensitively on the nonmagnetic impurities, but that the FFLO state itself is robust against disorder and survives even in the presence of the large impurity scattering, which is contrary to the general belief on the FFLO state\cite{D.F.Agterberg_JPCM_2001,A.B.Vorontsov_PRB_2008}.
Since the residual resistivity of CeCu$_2$Si$_2$ is much larger ($\rho_0 \sim 30~\mu \Omega$cm) than CeCoIn$_5$ ($\rho_0 \sim 3~\mu \Omega$cm), the difference in the phase diagram between CeCoIn$_5$ and CeCu$_2$Si$_2$ is ascribed to the the difference in the scattering rate by defects and/or disorder.       

It is noted that such an enhancement of $1/T_1T$ was not observed in UPd$_{2}$Al$_{3}$ which is also a candidate for observing the FFLO state.
We recently performed $^{27}$Al-NMR measurements on single-crystalline UPd$_{2}$Al$_{3}$ with the field parallel to the $c$ axis, and we discovered an unexpected symmetrical broadening in the NMR spectra in the field range 3~T $< \mu_0 H < \mu_0 H_{\rm c2} = 3.6$~T, suggesting the presence of a spatially inhomogeneous SC state such as the FFLO state\cite{S.Kitagawa_JPSJ_2017}.
However in the case of UPd$_{2}$Al$_{3}$, $1/T_1T$ did not increase, remaining at a constant value just below $\mu_0 H_{\rm c2}(T)$ before beginning to decrease below 3~T, in contrast to the results obtained using CeCu$_{2}$Si$_{2}$ and $\kappa$-(BEDT-TTF)$_{2}$Cu(NCS)$_{2}$.
Since the superconductivity in UPd$_2$Al$_3$ coexists with antiferromagnetic ordering, the FFLO state realized in UPd$_2$Al$_3$ is more complicated and may differ from that observed in CeCu$_{2}$Si$_{2}$ and $\kappa$-(BEDT-TTF)$_{2}$Cu(NCS)$_{2}$. 
A systematic NMR study near $\mu_0 H_{\rm c2}$ is needed in order to understand the characteristic features of the FFLO state, and it is currently in progress.

In conclusion, we have performed NMR measurements under a magnetic field close to $\mu_0 H_{\rm c2}$ in CeCu$_{2}$Si$_{2}$, which is a promising candidate for realizing the FFLO state.
In the SC state well below $\mu_0 H_{\rm c2}$, the Knight shift and $1/T_1T$ decreased abruptly below $T_{\rm c}(H)$, an effect that is well-understood within the framework of conventional spin-singlet superconductivity.
In contrast, $1/T_1T$ was enhanced just below $T_{\rm c}(H)$, exhibiting a broad maximum in the presence of a magnetic field close to $\mu_0H_{\rm c2}(0)$ applied parallel or perpendicular to the $c$ axis, although the Knight shift decreased just below $T_{\rm c}(H)$.
Comparing these results to those previously obtained for $\kappa$-(BEDT-TTF)$_{2}$Cu(NCS)$_{2}$, the enhancement of $1/T_1T$ implies the presence of the Andreev bound states in CeCu$_2$Si$_2$, which is regarded as a hallmark of the FFLO state.

\section*{Acknowledgments}
The authors acknowledge S. Yonezawa, Y. Maeno, Y. Tokiwa, Y. Yanase, and H. Ikeda for fruitful discussions. 
We would like to thank Editage for English language editing.
This work was partially supported by Kyoto Univ. LTM center, and Grants-in-Aid for Scientific Research (KAKENHI) (Grant Numbers JP15H05882, JP15H05884, JP15K21732, JP15H05745, and JP17K14339).

%

\end{document}